\documentclass[aps,prd,superscriptaddress,nofootinbib,groupedaddress,amsfonts,floatfix,twocolumn]{revtex4}
\usepackage{graphicx,amsmath,amssymb,amstext,amssymb,amsbsy,amsfonts,amsthm,lscape,units,tabularx,multirow}

\newcommand{\lp}{\left(}
\newcommand{\rp}{\right)}
\newcommand{\lb}{\left[}
\newcommand{\rb}{\right]}

\newcommand{\pr}{^{\prime}}
\newcommand{\prpr}{^{\prime \prime}}

\newcommand{\beq}{\begin{equation}}
\newcommand{\eeq}{\end{equation}}

\newcommand{\mch}{\mathcal{H}}

\newcommand{\eps}{\epsilon}

\newcommand{\delchi}{\Delta\chi}
\newcommand{\ttau}{\tilde{\tau}}
\newcommand{\btau}{\bar{\tau}}

\usepackage{color}

\begin{document}

\title{Dark Energy with a Triplet of Classical U(1) Fields}

\newcommand{\Dartmouth}{Department of Physics and Astronomy, Dartmouth College, 6127 Wilder Laboratory, Hanover, NH 03755}

\author{Avery~J.~Tishue}
\email{avery.tishue.gr@dartmouth.edu}
\author{Robert~R.~Caldwell}
\affiliation{\Dartmouth}

\begin{abstract}
We present a new mechanism for cosmic acceleration consisting of a scalar field coupled to a triplet of classical U(1) gauge fields. The gauge fields are arranged in a homogeneous, isotropic configuration, with both electric- and magnetic-like vacuum expectation values. The gauge fields provide a mass-like term via a Chern-Simons interaction that suspends the scalar away from its potential minimum, thereby enabling potential-dominated evolution. We show this mechanism can drive a brief period of acceleration, such as dark energy, without the need for fine tunings. We obtain simple analytic results for the dark energy equation of state and dependence on model parameters. In this model, the presence of the gauge field generically leads to a suppression of long-wavelength gravitational waves, with implications for the experimental search for cosmic microwave background B-modes and direct detection of a stochastic gravitational wave background.
\end{abstract}

\maketitle

\section{Introduction}
\label{sec:introduction}
Cosmic acceleration is a key pillar of the modern cosmological paradigm. An early inflationary epoch \cite{Guth:1980zm,Linde:1981mu,Albrecht:1982wi,Mukhanov:1981xt,Hawking:1982cz,Guth:1982ec,Bardeen:1983qw} is widely considered the leading approach for generating the initial conditions of the Hot Big Bang Cosmology, especially the quantum fluctuations that seed the observed large-scale structure and cosmic microwave background (CMB) anisotropies \cite{Planck:2018vyg}. Furthermore, evidence from type IA supernova (SNIA) and other probes indicates the late-time accelerated expansion of the Universe \cite{SupernovaSearchTeam:1998fmf,SupernovaCosmologyProject:1998vns}. Significant experimental effort has been devoted to understanding the fundamental nature of cosmic acceleration in both epochs. In the case of inflation, this takes the form of the search for primordial gravitational waves (GWs) \cite{Kamionkowski:2015yta,BICEPKeck:2022mhb,SimonsObservatory:2018koc,CMB-S4:2020lpa} and other observational consequences of inflation in the CMB \cite{Planck:2018jri}. In the case of late-time acceleration, the entire catalog of cosmological data is under scrutiny for some indication of the underlying physics, whether dark energy or other \cite{Planck:2015bue,Weinberg:2013agg}. At the same time, there has been concerted theoretical work to uncover and model the fundamental physical properties of the drivers of these accelerating epochs \cite{Martin:2013tda,Caldwell:2009ix,Copeland:2006wr}. Yet, despite this effort and progress, there remains no definitive model of either inflation or dark energy. 

A cosmological constant ($\Lambda$) provides a simple mechanism for cosmic acceleration. However, it fails as an inflationary candidate and offers no insight into the underlying physics of dark energy. Rather, the standard approach is to adopt an economy of mechanisms, and model each epoch of cosmic acceleration with a scalar field slowly rolling on a near-flat potential. Yet, it is a challenge to obtain a sufficiently flat potential from particle physics that leads to observationally viable inflation or dark energy. 

A new approach to drive accelerated expansion circumvents these challenges by coupling the scalar field to a gauge field via a Chern-Simons interaction. When the gauge field acquires a vacuum expectation value (vev), this coupling dynamically flattens the effective potential, thereby enabling slow roll acceleration on an otherwise steep potential. Scenarios based on this mechanism have been explored extensively in the literature, including realizations with U(1) \cite{Anber:2009ua,Fujita:2022fwc}, SU(2) \cite{Adshead:2016omu,Dimastrogiovanni:2016fuu,Adshead:2017hnc,Caldwell:2017chz,Domcke:2018rvv}, and SU(N) gauge fields \cite{Fujita:2021eue,Fujita:2022fff}, though other gauge groups are possible. In all cases, the background classical gauge field has isotropic, homogeneous stress-energy.  The advantage of these models is that they achieve slow roll dynamically, rather than through finely tuned parameters, designer potentials, or super-Planckian parameters and field excursions. Furthermore, these models exhibit interesting phenomenology, in part due to the parity-odd Chern-Simons scalar. Among the effects are GW-gauge field conversion \cite{Gertsenshteyn1962,Poznanin1969,Boccaletti1970,Zeldovich1974,Caldwell:2016sut,Fujita:2020rdx}, GW chirality, \cite{Anber:2012du,Adshead:2013nka,Adshead:2013qp,Maleknejad:2014wsa,Maleknejad:2016qjz}, GW opacity during the post-inflationary epoch \cite{Bielefeld:2014nza,Bielefeld:2015daa,Caldwell:2018feo,Tishue:2021blv}, non-Gaussianity \cite{Anber:2012du,Agrawal:2017awz,Dimastrogiovanni:2018xnn}, preheating \cite{Adshead:2015pva,Adshead:2017xll,Adshead:2018doq,Adshead:2019lbr}, magnetogenesis \cite{Dimopoulos:2001wx,Demozzi:2009fu,Kandus:2010nw,Adshead:2016iae}, and baryo- and lepto-genesis \cite{Noorbala:2012fh,Maleknejad:2014wsa,Maleknejad:2016dci,Caldwell:2017chz,Domcke:2019mnd}.

In this paper we propose a new mechanism for driving cosmic acceleration based on this approach. In particular, we study a toy model in which a scalar field $\chi$ is coupled to a triplet of classical U(1) gauge fields $A^I_\mu$ via a Chern-Simons interaction. The triplet of U(1) fields, set up in a flavor-space locked configuration, preserves homogeneity and isotropy at the background level and has both electric- and magnetic-like vevs. The Chern-Simons interaction, representing energy transfer between the scalar field and electric-like gauge vev, dynamically flattens the effective potential. Our main result, Eq.~(\ref{chic}), is an approximate analytic solution of the scalar field - gauge field system. We show that this model can accommodate a short burst of accelerated expansion, appropriate for a dark energy scenario. We obtain an analytic model of the equation of state, Eq.~(\ref{wchiE}), and discuss the observational consequences for the CMB and a spectrum of primordial GWs.

This paper is organized as follows. In Sec.~\ref{sec:model} we introduce the model and present analytic solutions.  In Sections ~\ref{sec:DEscenario} and \ref{sec:inflation} we apply this model to dark energy and inflationary scenarios, and discuss the theoretical considerations in each case. The implications of this model are discussed in Sec.~\ref{sec:discussion}. Calculation details are reserved for the Appendix.

\section{Model and Conventions}
\label{sec:model} 
We consider a toy model consisting of a scalar field, $\chi$, coupled  to a triplet of classical U(1) fields $A^{I}_{\mu}$ via a Chern-Simons interaction, where $I = 1-3$ indexes the U(1) fields.  The Lagrangian density for this system is 
\begin{align}
\mathcal{L} = &\frac{1}{2}M_P^2 R - \frac{1}{2}(\partial \chi)^2 - V(\chi) \nonumber\\ 
&- \frac{1}{4}F^{I}_{\mu \nu} F^{I \mu \nu} + \frac{\alpha}{8\pi}\frac{\chi}{f}F^{I}_{\mu \nu}\tilde{F}^{I \mu \nu} +\mathcal{L}_{rm}\label{lagrangian},
\end{align}
where the repeated index $I$ is summed over and $\mathcal{L}_{rm}$ is the Lagrangian density for the remaining matter and radiation. Throughout this paper we use the reduced Planck mass $M_P \equiv 1/\sqrt{8\pi G}$. We neglect the fermions associated with the gauge fields, assuming they are much heavier than the scales of interest. The field strength tensor for each U(1) field is $F^{I}_{\mu \nu} = \partial_{\mu}A^{I}_{\nu}-\partial_{\nu}A^{I}_{\mu}$, and the dual is $\tilde{F}^{I\mu \nu} = \epsilon^{\mu\nu\alpha\beta}F_{\alpha \beta}/2\sqrt{-g}$. Here $g=\rm{det} \, g_{\mu\nu}$, $\epsilon^{\mu\nu\alpha\beta}$ is the anti-symmetric Levi-Civita permutation symbol, and we take the convention $\epsilon^{0123}=1$. 

\subsection{Background and Ansatz}
\label{subsec:background}
We consider a homogeneous and isotropic Robertson-Walker (RW) background spacetime, with line element $ds^2 = a^{2}(\tau)(-d\tau^2 + d\mathbf{x}^2)$, where $\tau$ is the conformal time and the scale factor today is $a_0=1$. Varying the action with respect to the scalar and U(1) fields yields the equations of motion for each,
\begin{align}
    -\Box \chi + \frac{\partial V}{\partial \chi}  &=  \frac{\alpha}{8\pi f} F^{I}_{\mu \nu}\tilde{F}^{I\mu \nu} \label{scalareom}\\
   \nabla_{\sigma} F^{I\sigma \omega} &= \nabla_{\sigma} \left[\frac{\alpha \chi}{2\pi f}\tilde{F}^{I\sigma \omega} \right]. \label{U1eom}
\end{align}
Both the scalar field and gauge field equations of motion acquire new source terms due to the Chern-Simons interaction, which will alter the dynamics and serve to drive cosmic acceleration. 

To preserve the homogeneity and isotropy of the background cosmology, we assume $\chi = \chi(\tau)$. To the same end, the U(1) fields are set up in a flavor-space locked configuration:
\beq
F^{I}_{i0} = E(\tau) \delta^{I}_{i}, \quad F^{I}_{jk}=\epsilon^{I}_{\,jk}B(\tau), \label{flavorspacelock}
\eeq
such that the stress-energy tensor $T_{\mu \nu}^{I} = F^{I\alpha}_{\mu} F_{\nu \alpha}^{I} - g_{\mu\nu} (F^{I})^2/4$ is homogeneous and isotropic. Here $E(\tau)$ and $B(\tau)$ are the vevs of the gauge fields. With this flavor-space locked \textit{ansatz}, each U(1) copy is associated with a principal spatial direction and carries a parallel `electric' and `magnetic' field (which are not the fields of Standard Model electromagnetism). The electric and magnetic fields of each flavor must be parallel, or anti-parallel, otherwise they would generate a Poynting flux that would not respect the symmetry of the RW spacetime. Similarly, the electric (magnetic) vevs must be equal in amplitude for each flavor, otherwise isotropy would be violated. The isotropization of such configurations has been studied elsewhere, e.g. Ref.~\cite{Maleknejad:2012fw}.  

The energy density in the U(1) fields is $\rho_{\mathrm{U(1)}}=\sum_I \rho^I_{\mathrm{U(1)}}$, where $\rho_{\rm{U(1)}}^I = u^\mu u^\nu T^I_{\mu \nu}$ and $u^\mu$ is the four-velocity of an observer at rest with respect to the cosmic fluid. Likewise, the pressure in the $i = x,y,z$ direction is $p^{I\, i}_{\rm{U(1)}} = e^{i}_{\mu}e^{i}_{\nu} T^{I}_{\mu \nu}$ where $e^{i}_{\mu}$ is a set of three mutually orthogonal basis vectors. With the configuration in Eq.~(\ref{flavorspacelock}), the total energy density and pressure in the U(1) fields are
\begin{align}
    \rho_{\mathrm{U(1)}} &= \frac{3}{2a^4}(E^2 +B^2) = 3p_{\mathrm{U(1)}}.
\end{align}
The scalar field energy density has the standard form
\beq
\rho_{\chi} = \frac{1}{2}(\chi\pr/a)^2 + V(\chi).
\eeq
The equation of state of the coherent, classical U(1) fields is $w=1/3$, which means the negative pressure required for cosmic acceleration must be generated by the scalar field.

The gauge field equations of motion are as follows. From the Bianchi identity $\nabla_{[\mu}F_{\alpha \beta]} = 0$, we obtain Faraday's Law, $B\pr =0 $. Hence we take $B(\tau) = B_i$, a constant. The modified Amp\`ere-Maxwell equation becomes
\beq
E\pr = \frac{\alpha B_i}{2\pi f}\chi\pr,
\eeq
with solution 
\beq
E =E_i +\frac{\alpha B_i}{2\pi f}(\chi-\chi_i).
\label{eq:Evev}
\eeq
Here $E_i$ and $\chi_i$ are integration constants. The $E(\tau)$ vev is locked to the scalar field; even in the absence of an initial electric vev $E_i$, the scalar field causes one to develop. 

The scalar field - gauge field system is reduced to one effective degree of freedom, $\chi$. For simplicity we can assume $E_i = 0$, but our core proposal is independent of this assumption, which we later relax.  The equation of motion becomes
\beq
\chi\prpr + 2\mathcal{H}\chi\pr + a^2\frac{\partial V}{\partial \chi}  = -3a^{-2}\left(\frac{\alpha B_i}{2\pi f}\right)^2(\chi-\chi_{i}) \label{scalareomfull}.
\eeq
Heuristically, one can think of the term on the right hand side as arising from an effective mass term:
\begin{align}
V_{\mathrm{eff}} = V +\frac{1}{2}m_{CS}^2 (\chi-\chi_i)^2\\
m_{CS}^2 = \frac{3 }{a^4} \lp \frac{\alpha B_i}{2\pi f}\rp^2.
\end{align}
Now we can see how acceleration happens in this scenario. When $m_{CS}^2$ dominates the mass of the effective potential, it traps the scalar field near $V(\chi_i)$, leading to potential-dominated evolution.

\subsection{Driving Cosmic Acceleration}
\label{subsec:cosmicaccel}
We seek solutions in which $\chi$ is potential dominated as a consequence of the gauge interaction. For now, we will make the simplifying assumption of a quadratic potential, $V = \frac{1}{2} m^2 \chi^2$. Later, we envisage embedding this mechanism in a scenario where $\chi$ is taken to be the pseudo Nambu-Goldstone boson of a broken shift symmetry, and the potential has the form
$V = \mu^4 [1-\cos(\chi/f)]$.

With the quadratic potential, the scalar field equation of motion becomes
\beq
\chi\prpr + 2\mathcal{H}\chi\pr + a^2 \left[ m^2 \chi + 3 (a/a_i)^{-4}\left(\frac{ \hat{B}_i}{M}\right)^2 (\chi-\chi_i)\right] =0.\label{scalareomsimple}
\eeq
where $\hat{B}_i = B_i a_i^{-2}$ and $a_i$ is the scale factor at some arbitrary initial time when $E_i =0 $ and $\chi=\chi_i$.  We define $M \equiv 2\pi f / \alpha$ for notational compactness. The term in brackets gives the slope of the effective potential, $V_{\mathrm{eff}}\pr$. Acceleration occurs when $\chi$ slowly rolls in this effective potential, i.e. when the slope of the effective potential is small. This slope formally vanishes when $\chi$ is equal to a critical solution $\chi_c$:
\beq
\chi_c = \frac{\chi_i}{1+u} \label{chic},
\eeq
where we define $u_i \equiv ( mM / \sqrt{3} \hat{B}_i)^2$ and  $u\equiv u_i(a/a_i)^4$. Since $\chi_i>\chi_c$, $\chi$ inevitably rolls from $\chi_i$ to $\chi_c$.

The critical solution provides a good description of the evolution of the system under a wide range of conditions. To see this, we write $\chi(\tau) = \chi_c(\tau) + \delchi(\tau)$. The departure from the critical solution is described by the damped, driven harmonic oscillator equation
\begin{align}
\delchi\prpr + 2\mathcal{H}\delchi\pr + a^2 m^2 \left[1+u^{-1} \right]\delchi = F(a) \label{deltachiEOM}
\end{align}
where we define a driving term $F(a)$
\begin{align}
    F(a) \equiv 4 \chi_c \left[1-\frac{\chi_c}{\chi_i}\right] \left[\mch^2 \left(8-3\frac{\chi_{c}}{\chi_i}\right) + \frac{a\prpr}{a} \right]. \label{drivingterm}
\end{align}
Under the assumption $u \ll 1$, the initial amplitude is tiny, $\delchi_i \simeq u_i \chi_i$. As a result, the oscillations are negligible. The driving term is also tiny, owing to the $(1-\chi_c/\chi_i)$ prefactor. As a result, $\chi \simeq \chi_c$, as given in Eq.~(\ref{chic}), is an excellent description of the system. Additional details are provided in the Appendix.

A self-consistent potential-dominated solution is obtained as follows. Under the evolution given by the critical solution, the energy density in the $E$ vev can be written in terms of the scalar field potential energy, 
\beq
\rho_{E} = \frac{u}{(1 + u)^2} \frac{m^2 \chi_i^2}{2}.
\eeq
Next, $\rho_{\chi}$ and $\rho_{E}$ can be written as a single term,
\beq
\rho_{\chi E} \equiv \rho_{\chi}+\rho_{E} =  \frac{1}{1+u}\frac{m^2 \chi_i^2}{2}, \label{rhochiE}
\eeq
where we have assumed the scalar field kinetic energy is subdominant. The resulting equation of state of the scalar field - $E$ vev system is 
\beq
w_{\chi E}  = -1 + \frac{4 u}{3(1+u)}.\label{wchiE}
\eeq
We can see that $u \ll 1$ will yield an equation of state close to $-1$. By contrast, as $u$ approaches $1$, $w$ approaches $-1/3$, marking the end of acceleration.

The kinetic energy of the scalar field is subdominant when $R_{\chi_c}\ll 1 $ where
\beq
R_{\chi_c} \equiv \lp \frac{\chi_c^{\prime 2}/2a^2}{m^2\chi_c^2/2}\right) = \lp \frac{4u}{1+u} \frac{H}{m}\rp^2.
\eeq
Under reasonable assumptions about the background expansion, $H$ decays no faster than $a^{-4}$. The ratio $R_{\chi_c}$ grows while $u <1$ and decays once $u \gtrsim 1$, when acceleration has ended. Hence, the condition $R_{\chi_c} \ll 1$ is strictest as $u$ approaches unity, yielding a lower bound on $m$:
\beq
m \gg 2H|_{u=1}.
\eeq
A slightly stronger condition,
\beq
m \gg \sqrt{12}H|_{u=1}, \label{eqn:mHKEcond}
\eeq
ensures that the kinetic contributions to both the energy density and total equation of state are also subdominant to those of the $E$ vev. We also require that $\rho_{\chi E}$ dominates the total energy density. Together, these conditions can be used to select parameters for an accelerating scenario. 

Though we have taken $E_i=0$ for simplicity,
we note that this mechanism can accommodate a non-zero $E_i$. In this case, the critical solution has a modified amplitude compared to the $E_i=0$ case:
\beq
\chi_c =\frac{\chi_i}{1+u}\lp 1 - \frac{E_i M}{B_i \chi_i} \rp. \label{eq:chicEi}
\eeq
We must have $E_i B_i >0$ and $|E_i M /B_i \chi_i| \ll 1$; combined, these conditions ensure $\chi$ need not roll very far down from $\chi_i$ to reach the critical solution. Hence, the critical solution is only marginally modified compared the $E_i=0$ case. Using Eq.~(\ref{eq:chicEi}) in place of (\ref{chic}) in (\ref{eq:Evev}) does not change the form of the $E$ vev,
\beq
E = B_i\frac{\chi_c}{M}  \lp   \frac{\delchi}{\chi_c} -u\rp,
\eeq
although the evolution of $\delchi$ differs. Nevertheless, for a wide range of parameters $\delchi$ remains negligible compared to the critical solution. Cosmic acceleration proceeds similarly to the $E_i=0$ case, and the predictions are effectively the same. See Sec.~\ref{subsubsec:Ei} for details.

\section{Dark Energy Scenario}
\label{sec:DEscenario}
We begin with a brief overview of how the proposed mechanism can drive late-time acceleration. The system is initialized deep in the radiation era, for which $B=B_i$, $E_i = 0$, $\chi=\chi_i$, $\chi_i \pr =0$, and $u_i,\, u \ll 1$ at $a=a_i$. The scalar field evolves along the critical trajectory, $\chi_c$, and along with the $E$ vev has energy density $\rho_{DE} = \rho_{\chi E} = \rho_{\chi_i}/(1+u)$ and equation of state $w_{DE} = -1 + 4u/3(1+u)$. The trajectory of $(d\ln (1+w_{DE})/d \ln a,\, w_{DE})$ describes a thawing quintessence scenario, with $w_{DE}$ starting near $-1$ and slowly growing more positive \cite{Caldwell:2005tm}.  Oscillations and growth of $\delchi$ contribute negligibly to the energy density, as shown in Sec.~\ref{app:DEdeviations}. Initially the dark energy is subdominant, but as matter and radiation dilute, $\rho_{\chi E}$ comes to dominate and drive late-time acceleration, with $u \ll 1$ throughout.  For a suitable family of parameter choices $(m,M,B_i)$, the model yields viable present day values for $w_{DE,0}$ and $\Omega_{DE,0}$. At some $a>a_0$ in the future, $u$ approaches unity, and cosmic acceleration ends.
 
To construct a viable dark energy scenario, we connect the cosmological parameters to the model parameters. Combining Eqs.~(\ref{rhochiE}) and (\ref{wchiE}) relates $m$ and $\chi_i$ to $\Omega_{DE,0}$ and $w_{DE,0}$, 
\beq
\frac{m^2}{H_0^2}\frac{\chi_i^2}{M_P^2} = \frac{24\Omega_{DE,0}}{1-3w_{DE,0}}. \label{mH0req}
\eeq
Next, the constraint from the critical solution kinetic energy, $m \gg \sqrt{12}H_0$, yields an upper bound on $\chi_i/M_P$,
\beq
\frac{\chi_i}{M_P} \ll \sqrt{\frac{2 \Omega_{DE,0}}{1-3w_{DE,0}} }.
\eeq
In addition to the dark energy component, the $B$ vev contributes an energy density $\rho_B = 3 \hat{B}_i^2/2(a/a_i)^4$, which we parameterize as a fraction of the standard model cosmological radiation, $R_B=\rho_B / \rho_R$. Writing $u_i$ in terms of $R_B$, 
\beq
u_i   = \frac{1}{6 R_B \Omega_{R,0}}\left(\frac{m}{H_0}\right)^2 \left(\frac{M}{M_P}\right)^2 \lp\frac{a_i}{a_0}\rp^4,
\eeq
and using Eqs. (\ref{wchiE}) and (\ref{mH0req}), we obtain
\beq
 \lp\frac{M}{M_P} \rp^2 = \frac{3}{4} (1+w_{DE,0})\frac{R_B \Omega_{R,0}}{\Omega_{DE,0}}\lp \frac{\chi_i}{M_P}\rp^2\label{MMPreq}.
\eeq
With Eqs.~(\ref{mH0req}) and (\ref{MMPreq}) we can construct a fiducial model to drive late-time acceleration. For a specific realization, we set $\Omega_{DE,0} = 0.7$ and $w_{DE,0} = -0.95$, which are consistent with current observations \cite{Planck:2018vyg,Ross:2014qpa,DES:2021wwk,Brout:2022vxf,Jones:2022mvo,2011MNRAS.416.3017B,BOSS:2016wmc}. The ratio $R_B$ is constrained by considerations of Big Bang Nucleosynthesis (BBN) as well as the CMB, which constrain the presence of additional relativistic energy during the appropriate epoch. The bound is expressed in terms of the additional effective number of standard model neutrino species, $\Delta N_{\rm eff}$, beyond the Standard Model value $N_{\rm{eff}} = 3.046$,
\begin{equation}
    R_{B} \le \frac{\frac{7}{8}\left(\frac{4}{11}\right)^{4/3}\Delta N_{\rm eff} }{1 + \frac{7}{8}\left(\frac{4}{11}\right)^{4/3}N_{\rm eff} }.
\end{equation}
A joint BBN-CMB analysis gives $\Delta N_{\rm eff} < 0.168$~(95\% CL) \cite{Fields:2019pfx,Planck:2018vyg}, which translates to $R_B \lesssim 0.022$. To be conservative we take $\chi_i$ to be well below the Planck scale and $R_B$ to be well below the BBN-CMB bound: 
\beq
\left(R_B,\frac{\chi_i}{M_P} \right) = (2\times 10^{-3}, 10^{-5}).
\eeq
We note that values of $R_B$ close to the BBN-CMB limit help preserve the longevity of the critical solution.
This mechanism can then drive dark energy for 
\beq
\left(\frac{m}{H_0}, \frac{M}{M_P} \right) \simeq ( 2 \times 10^{5},  10^{-10} )
\eeq
where $u_i \simeq 4 \times 10^{-4} (a_i/a_0)^4$. 
For given values of $(w_{DE,0},\Omega_{DE,0})$ we have a three-parameter family of solutions for $R_B,\, m,\, M$.

We have not included $E_i$ in this analysis. The presence of an initial $E$ vev does not change the mechanism of acceleration. Moreover the initial value of $E_i$ is highly constrained, if we assume that the gauge vev never dominates cosmic expansion. As a consequence, $B_i \gg E$ is generic. (See Sec.~\ref{subsubsec:Ei}.)

\section{Inflationary Scenario?}
\label{sec:inflation}

An epoch of primordial acceleration is feasible in this model, but the parameters must be pushed to new extremes. To see this most clearly, we start with the requirement $u \ll 1$ throughout inflation, which serves to define the range of viable parameter space in terms of the number of efoldings. In particular, this yields the condition 
\beq
u_i = \left(\frac{M_P}{\chi_i} \right)^{2} \left(\frac{m^2 \chi_i^2}{3 \hat{B}_i^2}\right) \left( \frac{M}{M_P} \right)^2  \lesssim e^{-4N}. 
\label{pinfeqn}
\eeq
We require the first two terms in parenthesis to be much greater than unity, for sub-Planckian field strength, and for the scalar field potential to dominate the B vev energy density. As a result, $M$ must be exceptionally small. For a fiducial model, we select $\chi_i/M_P = 10^{-2}$ and $3 \hat{B}_i^2/m^2 \chi_i^2 = 10^{-2}$. We are free to adjust $m$ to obtain a suitable energy scale of inflation, $H_{inf}$. However, Eq.~(\ref{pinfeqn}) places an extraordinary demand on the Chern-Simons interaction coupling: for $N \simeq 60$ efoldings, we require $M \simeq 10^{-36}$~GeV, or $10^6 H_0$. The disparity in scales for the parameters is problematic for a standard inflationary epoch under this mechanism, as we now explain.

It is important to consider if such a small $M$ is consistently achievable in particle physics. For example, if the detailed origin of $\chi$ does not invoke a shift symmetry, this toy model simply contains a massive scalar coupled to the U(1) triplet through a Chern-Simons interaction. In this scenario, the small value of $M$, particularly in contrast to $m$, is not noteworthy. The mechanism described above can then drive inflation with parameters given by Eq.~(\ref{pinfeqn}), though the lack of shift symmetry potentially enables new interactions that can spoil the scenario.

Alternatively, $\chi$ may be the pseudo Nambu-Goldstone boson of a spontaneously broken global symmetry. In this case, $\chi$ has a residual shift symmetry $\chi \rightarrow \chi+2\pi n M$ for integer $n$, and the  potential $V(\chi)$ must respect this symmetry \cite{Marsh:2015xka}. In the simplest case, $V(\chi) = \mu^4 [1-\cos(\chi/M)]$. At the same time, the global symmetry has a quantum anomaly, which gives rise to a new term in the Lagrangian, the Chern-Simons interaction $(\chi/M) F \tilde{F}$. The $M$ appearing in the potential and the $M$ appearing in the Chern-Simons term are the same because both are the consequence of the same spontaneously broken global symmetry. In this scenario, $\mu^4 = m^2 M^2$. By examining Eq.~(\ref{pinfeqn}), it is easy to see $u_i \simeq e^{-4N}$ and $\rho_{\chi_i}> \rho_{B_i}$ are incompatible. There is no such issue for dark energy, since $\rho_{\chi_i} < \rho_{B_i}$. One possible workaround is that the $M$ appearing in the Chern-Simons term is exponentially suppressed compared to the $M$ appearing in the periodic potential, or more generally, that the Chern-Simons coupling is enhanced. Several mechanisms for achieving this have been proposed, although the requisite suppression of $M$ in the model presented here may be out of reach \cite{Kim:2004rp,Dimopoulos:2005ac,Bachlechner:2014hsa,Agrawal:2018mkd}.


\section{Discussion}
\label{sec:discussion}
In this work we have presented a new mechanism for driving late-time cosmic acceleration in which a scalar field is coupled through a Chern-Simons interaction to a triplet of classical U(1) gauge fields. The triplet is set up in a flavor-space locked configuration that preserves homogeneous and isotropic stress energy. The key ingredient is the presence of both magnetic- and electric-like gauge vevs, $B_i$ and $E(\tau)$. The modified Maxwell equations mean that any $B_i$ and dynamical $\chi$ cause the growth of the electric-like gauge vev. The scalar field, in turn, rolls along a modified effective potential due to the presence of the gauge vevs. As shown, this modification dynamically flattens the effective potential, permitting a new, slowly rolling critical solution for the scalar field, which can drive cosmic acceleration.

This model presents several interesting phenomenological possibilities for dark energy. Given the validity of the critical solution $\chi_c$, then for scenarios in which the equation of state remains close to $w_{\chi E}\simeq -1$, we have shown that the potential energy density of the critical solution, and energy density of the gauge field, dominate the $\chi-E$ system with negligible corrections. Rapid oscillations of $\delchi$ contribute a subdominant, dark matter-like component. Consequently, we expect the imprint on the CMB and matter power spectrum will closely resemble that of a $\Lambda$CDM model supplemented by a dark radiation component. That is to say, such a model is viable, in accord with current observations \cite{Aghanim:2018eyx}.

\begin{figure}[h]
\includegraphics[width=1\columnwidth]{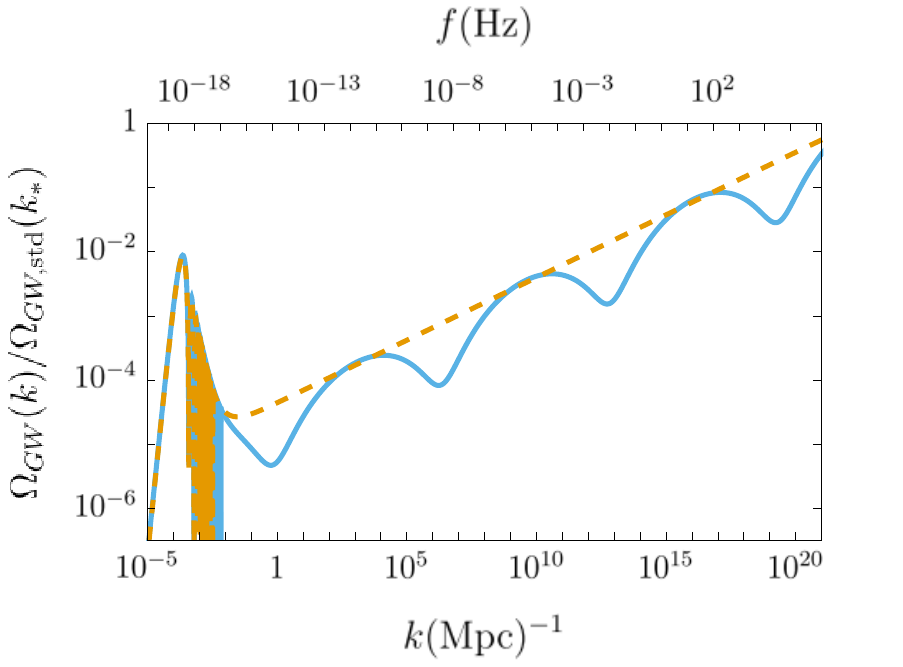}
\caption{Modification of stochastic GW background due to the presence of the gauge vevs. The dominant $B_i$ vev superimposes oscillations and a blue tilt onto an otherwise flat primordial spectrum, as shown in Ref.~\cite{Tishue:2021blv}. The full spectrum (light blue, solid) has a tilt that is well approximated using $n_T =  1- \sqrt{1-16R_B}$ (orange, dashed). For illustrative purposes, we have chosen the maximal value $R_B =0.022$; for smaller values the effect is reduced. The spectral density is normalized to a scale-invariant spectrum at $k_{*}=0.05$~$\mathrm{Mpc}^{-1}$.  }
\label{fig:GW}
\end{figure}

A unique imprint of this dark energy model is the transformation of a primordial GW background via GW-gauge field conversion. As we have shown elsewhere \cite{Tishue:2021blv}, the presence of a $B$-dominated gauge field during the radiation epoch acts to suppress super-horizon GW modes. The net result, illustrated in Fig.~\ref{fig:GW}, consists of a blue tilt $n_T= 1- \sqrt{1-16R_B}$ for modes $k > k_{eq}$, on top of an oscillatory modulation with logarithmic frequency, $\propto \cos^2\lb\sqrt{2R_B}\ln(k)\rb$ \cite{Tishue:2021blv}, superimposed on an otherwise scale-invariant spectrum. In this scenario, $B$-domination is generic; $E$-domination is highly constrained by cosmic history. Should current or future CMB experiments \cite{SimonsObservatory:2018koc,CMB-S4:2020lpa} detect a primordial B-mode pattern originating from a spectrum of primordial GWs, then there may be a novel, accompanying signal at high frequencies that is within reach of direct detection by future GW observatories \cite{Audley:2017drz}. In addition, we note that this phenomenon may ameliorate the overproduction of low frequency GWs in chromo-natural inflationary models \cite{Namba:2013kia,Adshead:2013qp}. Future work will examine the full imprint of fluctuations on the CMB and large scale structure.

\vspace{0.5cm}
\acknowledgments

This work is supported in part by U.S. Department of Energy Award No. DE-SC0010386. 
\vfill

\begin{widetext}
\noindent

\appendix

\section{Deviation from the Critical Solution \label{app:deviation}}
In this appendix we demonstrate that the deviation $\delchi$ from the critical solution contributes negligibly to the dynamics of the system and hence does not spoil cosmic acceleration and can be neglected. Specifically, we show that the contribution $\delchi$ makes to the energy density, equation of state, and $\chi=\chi_c+\delchi$ is insignificant. We will assume $E_i=0$ and handle the non-zero $E_i$ case in Section \ref{subsubsec:Ei}. Explicitly, under the decomposition $\chi= \chi_c +\delchi$, the total energy density in the scalar field and $E$ vev can be written 
\beq
\rho_{\chi E} = \frac{1}{2}m^2 \chi_c^2 \lb 1+u +\frac{(\delchi/\chi_c)^2}{u} + (\delchi/\chi_c)^2+ \lp\frac{4u H}{(1+u)m} \rp^2 \rb   + \frac{\delchi^{\prime 2}}{2a^2}+ \frac{\chi_c\pr \delchi\pr}{a^2} \label{rhodecomp}.
\eeq
The second and third terms in the brackets stem from the $\chi=\chi_c+\delchi$ decomposition in the $E$ vev, and we refer to these terms as $\rho_{E \chi_c}$ and $\rho_{E \delchi}$ respectively. The fourth term indicates the correction $\delchi$ makes to the potential energy, which we denote $\rho_{V \delchi}$---note the first order term cancels with the one stemming from the $E$ vev. The last term in brackets comes from the kinetic $\chi_c^{\prime 2}$ term, which the condition in Eq. (\ref{eqn:mHKEcond}) ensures can be dropped relative to the first two terms. Similarly, the $\chi_c\pr \delchi\pr$ cross term can be dropped because it is either subdominant to the $\delchi^{\prime 2}$ term or subdominant to the dropped $\chi_c^{\prime 2}$ term. Hence the energy density contributions stemming from $\delchi$ are given by $\rho_{\delchi} = \rho_{V \chi_c}\lb (\delchi/\chi_c)^2(1+u^{-1})\rb + \delchi^{\prime 2}/2a^2$, where because $u\ll1$ the $u$ term dominates and $\chi_c \approx \chi_i$.

This indicates that for $V_{\chi_c}$ to remain dominant requires $(\delchi/\chi_i)^2 \ll u$ and $\delchi^{\prime 2}/a^2m^2 \chi_i^2 \ll 1$. If instead $\delchi/\chi_i \ll  u$ and $\delchi^{\prime 2}/a^2m^2 \chi_i^2 \ll  u$, then the $\delchi$ contribution to the energy density (and equation of state) is also subdominant to that of the $\rho_{E \chi_c}$. To be concrete we set the bound that the $\delchi/\chi_i$, $\delchi\pr$, and $\chi_c\pr$ contributions to the energy density must remain smaller than $\delta = 10^{-5}$ that of the critical solution potential energy. The initial conditions for $\delchi$ automatically respect this because they give
\begin{align}
    \frac{\delchi_i}{\chi_i} &= u_i \\ 
    \frac{\delchi^{\prime \, 2}_{\, i}/a_i^2}{m^2 \chi_i^2} &=  \lp 4u_i \frac{H_i}{m}\rp^2,
\end{align}
where $u_i \ll 1$ (for dark energy $u_i =3(1+w_{DE,0}) (a_i/a_0)^4 /(1-3w_{DE,0}) $ while for inflation $u_i \simeq e^{-4N}$) and Eq.~(\ref{eqn:mHKEcond}) ensures the kinetic term is small, even compared to the $E$ vev. Hence $\rho_{\delchi}$ is initially negligible so we need only consider contributions to it that grow. We will find that these contributions grow monotonically. Hence evaluating Eq.~(\ref{rhodecomp}) for small $\delchi/\chi_i$ as $u \rightarrow 1$, we require
\begin{align}
\frac{H^2(u =1)}{m^2} &<  \frac{\delta}{2} \label{appHbound}\\
\left(\frac{\delchi}{\chi_i} \right)^2 &< \frac{\delta}{4}\label{appdelchibound}\\
\frac{\delchi^{\prime 2}/a^2}{m^2 \chi_i^2} & < \frac{ \delta}{2}\label{appdelchiprimebound},
\end{align}
Under these conditions, $V_{\chi_c}$ dominates and $\chi_c\pr$ and $\delchi$ can be safely ignored. Furthermore, as long as both $\delchi/\chi_i$ and $\delchi^{\prime 2}/a^2m^2 \chi_i^2$ remain smaller than $u$, then $\delchi$ contributes negligibly compared to the $E$ vev as well. 

To demonstrate that $\delchi$ obeys these bounds, we solve for its evolution, which is governed by
\begin{align}
\delchi\prpr + 2\mathcal{H}\delchi\pr + a^2 m^2 \left[1+ u^{-1} \right]\delchi = 4 \chi_c \left[1-\frac{\chi_c}{\chi_i}\right] \left[\mch^2 \left(8-3\frac{\chi_{c}}{\chi_i}\right) + \frac{a\prpr}{a} \right] \equiv F. \label{delchiEOMapp}
\end{align}
For $u \ll 1$ the driving term, $F$, can be approximated
\begin{align}
F&\simeq 20 \delchi_i (a/a_i)^4   \mch^2\left[ 1 +\frac{1}{10}(1-3w) \right].
\end{align}
We approximate the evolution of $\delchi$ by specifying the background evolution exactly. For inflation, we take $w = -1$ and we show $\rho_{\delchi}$ strictly decays. For dark energy, we assume a piecewise continuous cosmology with successive radiation- and matter-dominated backgrounds. We are only concerned with the evolution until today, which is not fully dark energy dominated, so we will take matter domination to last until $a_0$ and not consider a dark energy dominated background. This approximation is reasonable because the period of dark energy dominated background evolution is not very long and hence makes small corrections to the evolution of $\delchi$, and furthermore $\rho_{\delchi}$ decays in a $w=-1$ background.
\subsection{Dark Energy \label{app:DEdeviations} }
\subsubsection{Radiation Epoch}
For a radiation dominated background, $F\simeq 20  \delchi_i u^4 \mch^2
\simeq 20  \delchi_i  \tau^2/\tau_i^4$. The subscript $``i"$ corresponds to the initial conditions, which for dark energy we choose to be deep in the radiation epoch for which $H_i^2 \simeq H_0^2 (a_0/a_i)^4 \Omega_{R,0}(1+R_B)$. The general solution in the radiation epoch is 
\begin{align}
\delchi/\delchi_i = \frac{20\bar{\tau}^4}{20+\kappa} + A_+ \bar{\tau}^{-\frac{1}{2}(1+\sqrt{1-4\kappa})} + A_- \bar{\tau}^{-\frac{1}{2}(1-\sqrt{1-4\kappa})} \label{raddevgeneralsolution}
\end{align}
where $\btau \equiv \tau/\tau_i$ and $\kappa \equiv m^2 a_i^2 \tau_i^2/ u_i$. Imposing a radiation background and the conditions for a viable background model, Eqs. (\ref{wchiE}) and (\ref{mH0req}),  we can rewrite $\kappa$
\beq
\kappa = 8 \frac{\Omega_{DE,0}}{\Omega_{R,0}(1+R_B)} \frac{(\chi_i/M_P)^{-2}}{1+w_{DE,0}}. \label{kappavalue}
\eeq
This is much larger than unity because $w_{\chi,0} \lesssim -0.9$ and we require $\chi_i/M_P < 1$.  This means Eq.~(\ref{kappavalue}) gives $\kappa a_{eq} \gtrsim 177 $ (neglecting the small contribution of $R_B$), and more realistically $\kappa a_{eq} \gtrsim 10^4 $.  This inequality becomes more severe for smaller $\chi_i$ and more negative equation of state $w_{DE,0}$. For $\kappa \gg 1$ we have $\sqrt{1-4\kappa} \simeq 2i\sqrt{\kappa}$. In this case the solution Eq.~(\ref{raddevgeneralsolution}) obtains oscillating terms, which we take the real part of, yielding
\begin{align}
\delchi / \delchi_i = \frac{20}{20+\kappa}\btau^4 + B_+ \btau^{-1/2} \cos(\kappa^{1/2}\ln(\btau)) +  B_- \btau^{-1/2} \sin(\kappa^{1/2}\ln(\btau)) \label{DEdevradsoln2} .
\end{align}

The initial conditions give the coefficients as
\begin{align}
    B_+ &= 1- \frac{20}{20+\kappa} \\
    B_- &= \frac{5}{2}\kappa^{-1/2} \left(1 - \frac{36}{20+\kappa} \right),
\end{align}
so $B_+ \gg B_-$. The solution Eq.~(\ref{DEdevradsoln2}) has a growing term and decaying oscillating terms. When computing $\delchi^2$ and $\delchi^{\prime 2}$, the oscillating terms on their own can only contribute decaying, and hence negligible, potential and kinetic energy. The cross term is either no bigger than the squared decaying term, or no bigger than the squared growing term, hence we need only consider
\beq
\frac{\delchi}{\chi_i} \simeq \frac{20}{20+\kappa} u \ll u,
\eeq
so $\delchi$ is negligible compared to the $E$ vev, and is largest at $a=a_{eq}$, for which it is well below the designated bound $(\delchi/\chi_i) < (10^{-5}/4)^{1/2}$. The kinetic contribution gives
\begin{align}
    \frac{\delchi^{\prime 2}}{a^2 m^2 \chi_i^2}  = \lp \frac{80}{20+\kappa}\rp^2 \frac{ u}{\kappa}.
\end{align}
This is also smaller than $u$ and largest for $a=a_{eq}$, for which it is well below the designated $10^{-5}$ bound. Thus during the radiation era $\delchi$ is never significant. We demonstrate the validity of these analytic results as well as the subdominance of the $\delchi$ contributions to the energy density in Figures \ref{fig:radanalytics} and \ref{fig:radrho}.

\begin{figure}[h!]
\includegraphics[width=1\textwidth]{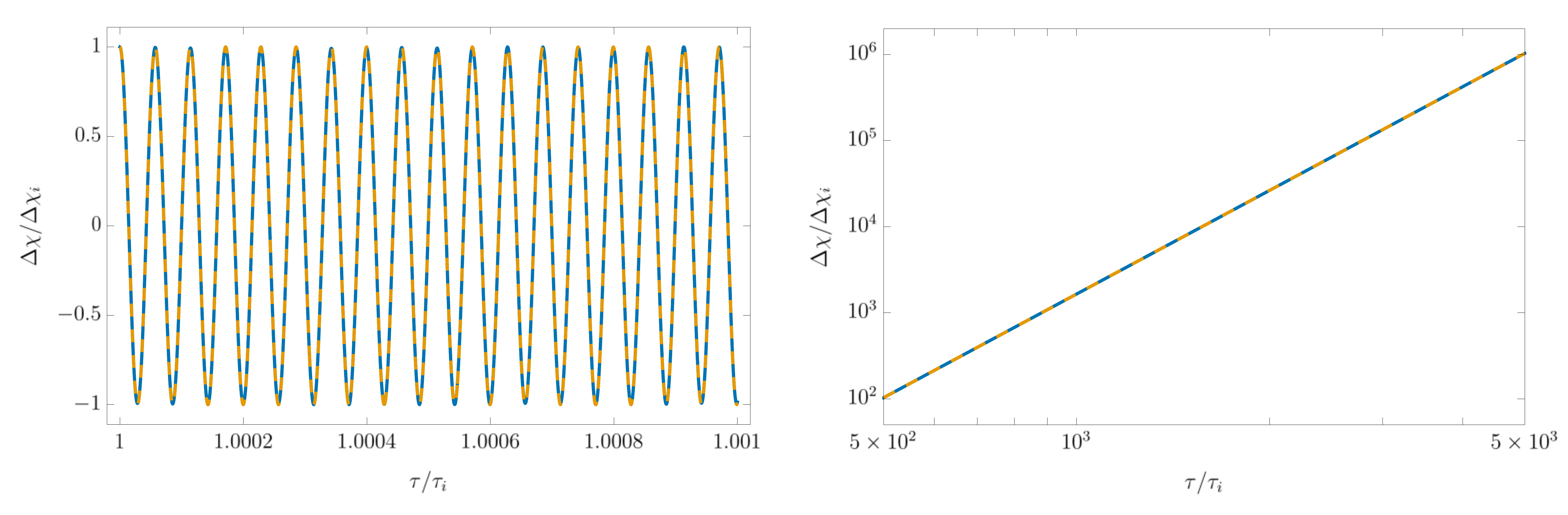}
\caption{Demonstrations of validity of radiation-era analytic solution (blue) of $\delchi$, Eq.~(\ref{raddevgeneralsolution}), compared to numerical solution (orange, dashed). The oscillating term dominates at early times (left panel) while at later times the $a^4$ growing behavior dominates (right panel). The evolution is given by a model with $(\chi_i/M_P,m/H_0,M/M_P,R_B)=(10^{-2},250,2.2\times10^{-6},0.01)$, which yields $(\Omega_{DE,0},w_{DE,0})=(0.7,-0.95)$.}
\label{fig:radanalytics}
\end{figure}

\begin{figure}[h!]
\includegraphics[width=1\textwidth]{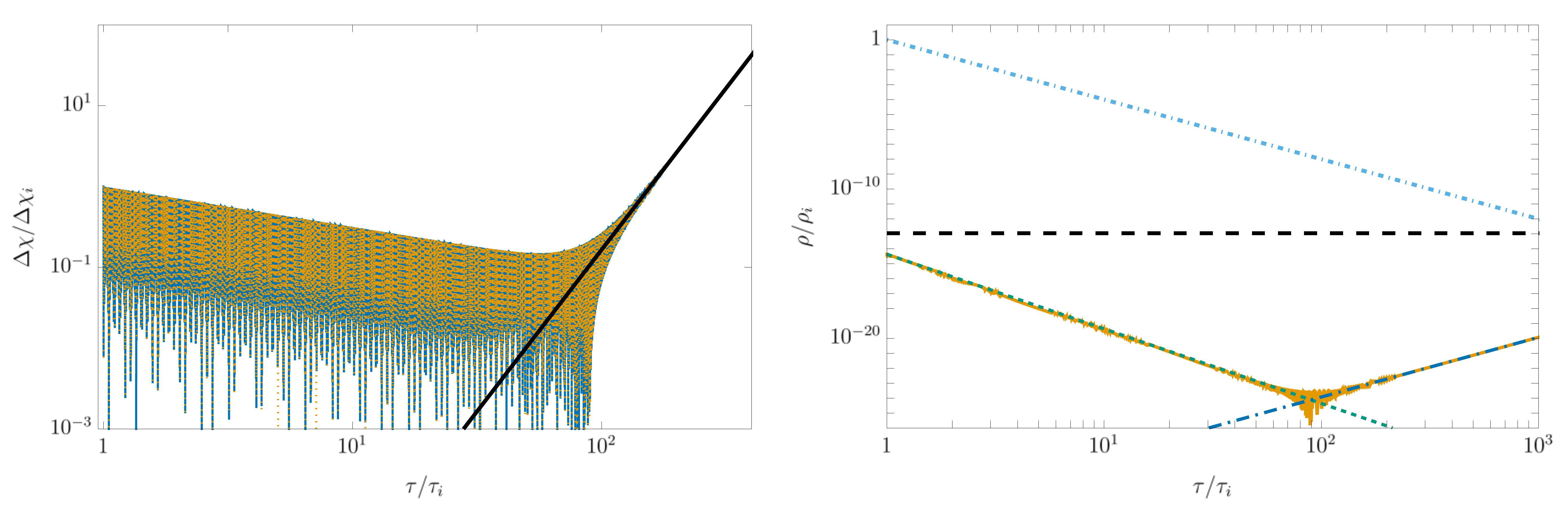}
\caption{Left panel: Full numerical (orange, dotted) and analytic (blue) solutions of $\delchi$ during radiation domination, demonstrating the transition from oscillation to strict $a^4$ growth (black, dashed). Right panel: Comparison of numerically computed energy densities, in units of initial energy density, of the total energy density (light blue, dot-dashed), $10^{83 }V_{\chi_c}$ (black, dashed), and  $10^{183} \rho_{\delchi}$ (orange, solid), where $\rho_{\delchi}$ includes all terms $\delchi$ contributes to the total $\rho$. The numerical scaling of $V_{\chi_c}$ and  $\rho_{\delchi}$ is arbitrary and is intended for visibility. Initially $\rho_{\delchi}$ decays as $a^{-5}$ (green,dotted), and later transitions to $a^4$ growth (dark blue, dot-long-dashed). The model parameters are the same as in Figure \ref{fig:radanalytics}.}
\label{fig:radrho}
\end{figure}

\subsubsection{Matter Epoch}
In the matter epoch, $F\simeq 22 \delchi_i u^4 \mch^2$, which yields the solution
\begin{align}
\delchi =& A_- \left[\frac{\kappa^{1/2}}{\ttau} \cos\left(\frac{\kappa^{1/2}}{\ttau} - \kappa^{1/2} \right) - \sin \left(\frac{\kappa^{1/2}}{\ttau} - \kappa^{1/2} \right) \right] \nonumber \\ 
& + A_+ \left[\cos\left(\frac{\kappa^{1/2}}{\ttau} - \kappa^{1/2} \right) + \frac{\kappa^{1/2}}{\ttau}  \sin \left(\frac{\kappa^{1/2}}{\ttau} - \kappa^{1/2} \right) \right] + f(\kappa,\tilde{\tau})
\end{align}
where $\tilde{\tau} \equiv \tau/\tau_{eq}$, $f$ is given by 
\begin{align}
f(\kappa,\tilde{\tau}) =& 88 \delchi_i (a_{eq}/a_i)^4 \frac{\tilde{\tau}^{10}}{\kappa} \biggl[1 -5e^{-i\frac{\kappa^{1/2}}{\tilde{\tau}}}(1-i\frac{\kappa^{1/2}}{\tilde{\tau}})E_{12}(-i\frac{\kappa^{1/2}}{\tilde{\tau}})  -5e^{i\frac{\kappa^{1/2}}{\tilde{\tau}}}(1+i\frac{\kappa^{1/2}}{\tilde{\tau}})E_{12}(i\frac{\kappa^{1/2}}{\tilde{\tau}}) \biggr],
\end{align}
and the exponential integral is 
\beq
E_n(z) = \int_1^{\infty} \frac{e^{-zt}}{t^n}dt.
\eeq
The function $f$ is purely real, and because $\kappa^{1/2}/ \ttau \gtrsim 177^{1/2}$ (a simple consequence of Eq.~(\ref{kappavalue})), it is dominated by the first term in the brackets, i.e. to an excellent approximation we can take 
\beq
f(\kappa,\tilde{\tau}) = 88 \delchi_i (a_{eq}/a_i)^4 \frac{\tilde{\tau}^{10}}{\kappa}.
\eeq 
This approximation is off by a factor of $\sim 2$ for the limiting value of $\kappa^{1/2}/ \ttau \approx 177^{1/2}$, but quickly reaches $\sim 99\%$ accuracy for $\kappa^{1/2}/ \ttau = 100$, and $\sim 99.999\%$ accuracy for $\kappa^{1/2}/ \ttau = 3\times 10^3$. Ensuring these accuracies by $a_0$ requires $\kappa \gtrsim 10^8$ and $\kappa \gtrsim 10^{11}$ respectively, which from Eq.~(\ref{kappavalue}) translates to modest upper bounds on $\chi_i/M_P$ via Eq.~(\ref{kappavalue}).

In a piecewise cosmology approximation the coefficients $A_{\pm}$ are determined by matching $\delchi,\delchi\pr$ from the radiation and matter era solutions at $\tau_{eq}$,
\begin{align}
    A_+ = \kappa^{-1} \left[ 880 \left(a_{eq}/a_i\right)^{4} \delchi_i \kappa^{-1} - \tau_{eq} \delchi\pr_{eq} \right],
\end{align}
\begin{align}
    A_- =\, \kappa^{-1/2} \biggl[&\delchi_{eq}\left( 1 - 88 (a_{eq}/a_i)^4 \kappa^{-1} \frac{\delchi_i}{\delchi_{eq}}
    \right) +  \kappa^{-1} \left( \tau_{eq} \delchi\pr_{eq} - 880 (a_{eq}/a_i)^4  \kappa^{-1}\delchi_i \right)\biggr].
\end{align}
Inputting the radiation solution at equality, which is the pure growing mode, gives $|A_-| \gg |A_+|$, and also indicates that the growing $f$ term is initially slightly larger than the $A_{-}$ term. We need only growing contributions to $\rho_{\delchi}$ anyhow, and hence we can ignore the oscillating terms, which gives
\beq
\frac{\delchi}{\chi_i} = \frac{88 u}{\kappa} \frac{a}{a_{eq}}
\eeq
from which we have $\delchi/\chi_i <  u$. To respect the designated bound (\ref{appdelchibound}), we rewrite this using Eq.~(\ref{kappavalue}),
\beq
\frac{\delchi}{\chi_i} = 33 \frac{\Omega_{M,0}(1+R_B)}{\Omega_{DE,0}} \lp \frac{a}{a_0}\rp a \frac{(1+w_{DE,0})^2}{1-3w_{DE,0}} \lp \frac{\chi_i}{M_P} \rp^2.
\eeq
The bound on $\delchi/\chi_i$ thus translates to a bound on $\chi_i$, which is strictest for $a=a_0$ and large $w_{DE,0}$. Taking $w_{DE,0}=-0.9$ yields the modest constraint $\chi_i/M_P \lesssim 0.2$. The kinetic energy contribution gives
\beq
\lp \frac{\delchi\pr}{a m\chi_i} \rp^2 = \frac{880^2  u}{\kappa^3}(a/a_{eq})^3,
\eeq
which also is less than $ u$. This also yields a bound on $\chi_i/M_P$, 
\beq
(\chi_i/M_P)^6 = \lp \frac{\delchi\pr}{am\chi_i} \rp^2 \frac{9075}{2} \frac{(1+w_{DE,0})^4}{1-3w_{DE,0}} \lp \frac{\Omega_{M,0}}{\Omega_{DE,0}} \rp^3 a^3.
\eeq
The strictest bound gives $\chi_i/M_P \lesssim 0.31$. 
It is possible for one of the oscillating terms to dominate the kinetic term if $\kappa \gtrsim (220/17)^2(a_0/a_{eq})^{6}\approx 2.3 \times 10^{23}$ in which case the kinetic term is bounded from above,
\beq
\lp \frac{\delchi\pr}{a m\chi_i} \rp^2 = \frac{68^2  u}{\kappa^2}(a_{eq}/a)^8 (a_0/a).
\eeq
This is maximized at $a_{eq}$ and is much smaller than u and decaying, and hence can be neglected. Hence as long as $\chi_i/M_P \lesssim 0.2$, $\delchi$ is negligible compared to both $\chi_c$ and the $E$ vev, and it does not spoil dark energy prematurely. 

This piecewise cosmology provides a good rough estimate that indicates $\delchi$ is negligible, but in practice the background cosmology does not undergo a prolonged period of $w=0$ expansion to high accuracy. To confirm our approximation we supplement these analytic estimates with numerical results, shown in Figure \ref{fig:matrho}, that demonstrate $\delchi$ remains negligible. In the example shown, the energy contribution is $\rho_{\delchi}(a_0)/\rho_0 \approx 1.4 \times 10^{-9}$.

\begin{figure}[h!]
\includegraphics[width=1\textwidth]{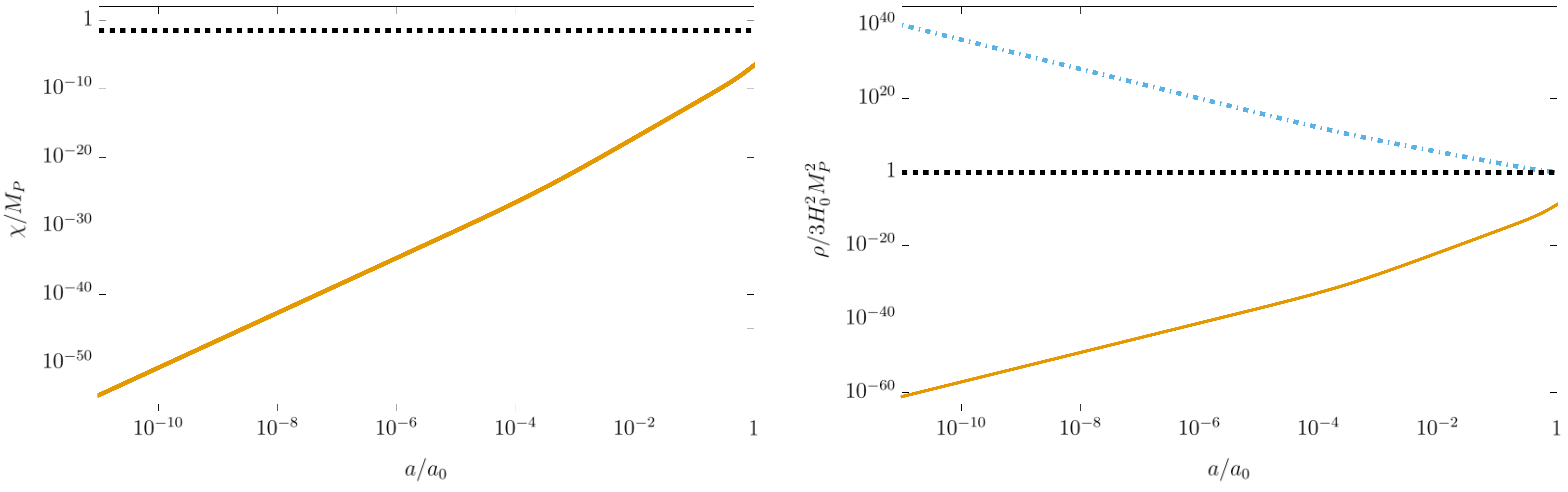}
\caption{Behavior of $\delchi$ in a model with $(\chi_i/M_P,m/H_0,M/M_P,R_B)=(10^{-1.5},66.1,9.95\times 10^{-6},0.02)$, which yields $(\Omega_{DE,0},w_{DE,0})=(0.7,-0.95)$. Left panel: Numerical evolution of $\chi_c$ (black, dotted) and $\delchi$ (orange, solid) from radiation domination to today. Right panel: Comparison of numerically computed energy densities, in units of $\rho_0$, of the total energy density (light blue, dot-dashed), $V_{\chi_c}$ (black, dotted), and  $\rho_{\delchi}$ (orange, solid).}
\label{fig:matrho}
\end{figure}

\subsubsection{Nonzero $E_i$}
\label{subsubsec:Ei}
As explained in Sec.~\ref{subsec:cosmicaccel}, the inclusion of a nonzero $E_i$ has the effect of introducing a small, constant modification to the critical solution,  $\chi_c=\chi_i(1-j)/(1+u)$ where $j \equiv E_i M / B_i \chi_i \ll 1$. The $E$ vev takes the same form as before, 
\beq
E= \frac{B_i \chi_i(1-j)}{M(1+u)}u \lb -1 + u^{-1} \frac{\delchi(1+u)}{\chi_i(1-j)} \rb = B_i\frac{\chi_c}{M}  u \lb u^{-1} \frac{\delchi}{\chi_c} -1\rb \label{EeqnEi}.
\eeq
The energy density in the scalar field and $E$ vev, $\rho_{\chi E}$, takes the same form as before, Eq.~(\ref{rhodecomp}), and $\chi_c$ is effectively unchanged. Hence our conditions (\ref{appHbound})-(\ref{appdelchiprimebound}) on the contributions from $\delchi$ are unchanged. 
However, the initial conditions and evolution of $\delchi$ are changed: $\delchi_i/\chi_i \approx u_i+j$ and $\delchi\pr_i/\chi_i \approx 4u_i \tau_i^{-1}(1-2u_i-j) \approx 4u_i \tau_i^{-1}$, and the equation of motion acquires an additional source term
\begin{align}
\delchi\prpr + 2\mathcal{H}\delchi\pr + \frac{m^2 a_i^2}{u_i}\frac{a_i^2}{a^2}\delchi \approx 20 \chi_i (u_i(a/a_i)^4 +j)  \mch^2\left[ 1 +\frac{1}{10}(1-3w) \right].
\end{align}
Note the initial kinetic energy contribution of the deviation is unchanged, but the larger $\delchi_i$ value means the initial potential energy contribution from $\delchi$ is larger, as is the initial energy density that $\delchi$ contributes via $E$, i.e. the $(\delchi/\chi_c)^2/u$ term. This fact, combined with the altered dynamics, means this scenario can be appreciably different than the $E_i=0$ case.

In the $E_i=0$ scenario, we showed the $\delchi$ contributions in $\rho_{\chi E}$ are all smaller than not only $\rho_{\chi_c}$ but also $\rho_{E \chi_c}$, and hence they do not spoil cosmic acceleration and their effects on the equation of state and energy density are totally subdominant to those of $\chi_c$ and $E$. While this is sufficient for a dark energy scenario, it is not necessary. In particular, the $\rho_{E \delchi}=V_{\chi_c} (\delchi/\chi_c)^2/u$ term can be larger than $V_{\chi_c}$ before dark energy domination as long as it is well below the energy density of the radiation or matter that dominates the background evolution. That is, we require $V_{\chi_c} (\delchi/\chi_c)^2/u \ll 3 \mch^2 a^2 M_P^2$. In both the radiation and matter dominated epochs, this leads to the same condition:
\beq
(\delchi/\chi_i)^2 \ll u_0 \frac{\Omega_{R,0}}{\Omega_{DE,0}}. \label{delchiureq}
\eeq
Note that while $\delchi$ does contribute to the total energy density and equation of state, we do not include the $\delchi$ contributions in the definitions of $\rho_{DE}$ and $w_{DE}$ because $\rho_{\delchi}$ does not evolve as a simple modulation of $V_{\chi_c}$ the way $\rho_{E \chi_c}$ does. To ensure dark energy via the critical solution is viable, we require the same conditions from before, Eqs. (\ref{appHbound})-(\ref{appdelchiprimebound}), which bound $V_{\delchi}$ and $\delchi^{\prime 2}/2 a^2$ compared to $V_{\chi_c}$.

In  a pure radiation background, the solution acquires an additional constant term,
\begin{align}
\delchi / \chi_i = \frac{20j }{\kappa} 
+ \frac{20u_i}{20+\kappa}\btau^4 + B_+ \btau^{-1/2} \cos(\kappa^{1/2}\ln(\btau)) +  B_- \btau^{-1/2} \sin(\kappa^{1/2}\ln(\btau)),
\label{delchiradEi}
\end{align}
where
\begin{align}
B_+ = j(1+20 \kappa^{-1}) + u_i \kappa/(20+ \kappa) \approx u_i+j \\
B_- = \frac{j(400+\kappa^2)+ 9 \kappa^2 u_i}{2 \kappa^{3/2}(20+\kappa)} \approx (9u_i+j)/(2 \kappa^{1/2}).
\end{align}
Note that the growing term remains the same as in the $E_i=0$ case. For $j<u_i$ we recover the $E_i=0$ behavior, and hence we need only study the $j> u_i $ case, and we have $\delchi_i \approx j$. Initially $\delchi$ decays, and if the growing term ever dominates, we are returned to the sufficiently bounded $E_i=0$ scenario. Hence we need only consider $\delchi_i/\chi_i$, which gives 
\begin{align}
j^2 &\ll \delta/4 \\
    j^2 &\ll u_0 \frac{\Omega_{R,0}}{\Omega_{DE,0}}
\end{align}
from Eqs. (\ref{appdelchibound}) and (\ref{delchiureq}) respectively. The kinetic contribution is similarly simple, as the form of $\delchi\pr$ is the same as in the $E_i=0$ case, but with modified $B_{\pm}$ coefficients. Hence either the kinetic energy has strictly decayed from $a_i$ to $a_{eq}$, or else the growing term dominates if $\btau_{kin}=[(j \kappa^{3/2}/20u_i)]^{2/9} > \btau_{eq}$, for which case the kinetic contribution is the same as in the $E_i=0$ case.

In a pure matter background, $\delchi$ acquires new terms comprised of special functions owing to the new source term; similar to the $E_i=0$ case, we can again study this in the $\sqrt{\kappa}>>\ttau$ limit for which the solution is
\begin{align}
\delchi/\chi_i \approx& \frac{88 u_i (a_{eq}/a_i)^4}{\kappa}\ttau^{10} + \frac{88 j}{\kappa} \ttau^2 +  A_- \left[\frac{\kappa^{1/2}}{\ttau} \cos\left(\frac{\kappa^{1/2}}{\ttau} - \kappa^{1/2} \right) - \sin \left(\frac{\kappa^{1/2}}{\ttau} - \kappa^{1/2} \right) \right] \nonumber \\ 
& + A_+ \left[\cos\left(\frac{\kappa^{1/2}}{\ttau} - \kappa^{1/2} \right) + \frac{\kappa^{1/2}}{\ttau}  \sin \left(\frac{\kappa^{1/2}}{\ttau} - \kappa^{1/2} \right) \right] . \label{Eimatterdelchi}
\end{align}
The effect of $E_i$ on $\delchi$ in the matter era is to add a new growing term $\propto j \ttau^2$, and alter the coefficients $A_{\pm}$ of the oscillating terms, while the $\ttau^{10}$ term is the same as in the $E_i=0$ case. Much of the same logic used in previous sections can be used here. The oscillating terms in $\delchi$ are at best constant on average, and decaying as $\ttau ^{-3}$ in $\delchi\pr$. Because we have shown the deviation is negligible at $a_{eq}$, we again need only consider the growing terms. We have already shown in the $E_i=0$ case that the $\ttau^{10}$ term is sufficiently bounded. Hence we need only consider the new term, which is largest today and gives the  potential energy bound 
\beq 
j \lesssim \frac{\kappa}{88} \frac{a_e}{a_0} (\delta/4)^{1/2}.
\eeq
Note this term automatically satisfies $(\delchi/\chi_i)<u$ because $\kappa u_0 \approx 6 (\chi_i/M_P)^{-2} (\Omega_{DE,0}/ \Omega_{R,0})$.
This term on its own generates a decaying energy density,
\beq
\lp \frac{\delchi\pr}{am \chi_i}\rp^2 = \frac{176 j^2}{\kappa^3 p(a_{eq}/a_i)^4} \frac{a_e}{a}
\eeq
and hence can be neglected. The $\ttau^{10}$ term already satisfies $(\delchi/\chi_i)^2 < u^2$, and to satisfy (\ref{delchiureq}) imposes the constraint
\beq
\kappa^2 \gg 88^2 u_0 \frac{\Omega_{DE,0} a_0^2}{\Omega_{R,0}a_{eq}^2},
\eeq
which is automatically satisfied by $\kappa = \kappa_{min}$. To satisfy the same constraint if the $j$ term dominates requires 
\beq
(j/\kappa)^2 \ll  \frac{u_0}{88^2} \frac{\Omega_{R,0}a_{eq}^2}{\Omega_{DE,0}a_0^2},
\eeq
which combined with the above condition yields the sufficient but not necessary condition $j<u_0$. The conditions laid out in this section ensure $\delchi$ does not spoil the cosmic acceleration driven by $\chi_c$. Note also that Eq.~(\ref{Eimatterdelchi}) easily satisfies $\delchi/\chi_i < u$ at $a_0$, and hence this case still recovers the late time behavior of $w_{\chi E}$ from the $E_i = 0$ case.

The behavior of $E(a)$ may also be significantly different than the $E_i=0$ case. This is important because the gauge vevs will modify GW evolution, and in particular the tilt superimposed on a flat GW spectrum depends on the relative amplitudes of the $E$ and $B$ vevs \cite{Tishue:2021blv}. A blue (red) tilt is superimposed if $B$ ($E$) dominates, leading to a suppressed (enhanced) GW spectrum at low frequencies. For the $E_i=0$ case, $\delchi/\chi_i < u $ throughout the radiation epoch, and so Eq.~(\ref{EeqnEi}) gives $E/B_i \simeq - (\chi_i/M)u$.  This gives $|E/B_i| \ll1$ for large regions of the parameter space (see Figure \ref{fig:RB_chiM}), and hence the GW spectrum is uniformly suppressed. However, the behavior is different for non-zero $E_i$. For this case, we need only consider the new term and dominant ($B_+$) decaying term in Eq.~(\ref{delchiradEi}), as the $\ttau^4$ growing term is the same as for $E_i=0$, i.e. it gives $\delchi/\chi_i \ll u$ and cannot give $|E/B_i|>1$. These two terms give
\beq
\delchi/\chi_i \simeq j \lb \frac{20}{\kappa} + \ttau^{-1/2} \cos(\kappa^{1/2}\ln(\ttau)) \rb,
\eeq
This can easily satisfy $\delchi/\chi_i > u$, and hence $E/B_i \simeq (\chi_i/M)(\delchi/\chi_i)$, for some or all of the radiation epoch. In turn, for sufficiently large $\chi_i/M$ and not too small $\delchi/\chi_i$, there can be an extended period of $|E/B_i|>1$. The biggest inequality comes for small $\kappa$, which is bounded by $\kappa_{min}\approx 10^7$, and for large $j$. As a limiting example, we can consider $j \approx 10^{-5}$ and $\kappa \approx \kappa_{min}$. In this scenario, $\delchi/\chi_i$ is roughly $10^{-11}$ by $a_{eq}$. At the same time, $\rho_{E}$ decays by at least $10^{12}$ relative to $\rho_B$ and $\rho_R$ throughout the radiation epoch. Therefore, because $\rho_E$ is initially small compared to the background (i.e. $R_{E,i} < 0.022$), it quickly becomes even smaller. This means the already modest $E$-enhancement of the GWs is further reduced, and even small $R_B$ fractions will cause low frequency GW modes to be net suppressed. The relationship between important model parameters is shown in Fig.~\ref{fig:RB_chiM}. The upshot is that for large swaths of parameter space this model generically predicts a $B$-dominated gauge vev, while $E$-dominated gauge vevs correspond to extremal models, e.g. those with $w_0$ exceptionally close to $-1$. Generically, the GW spectrum acquires a blue-tilt and a suppression at long wavelengths. The caveat to this is the novel case in which well before BBN, we permit the $E$ vev to dominate the background evolution even compared to standard model radiation --- we leave such a scenario and its implications for GW evolution for future study.

\begin{figure}
    \centering
    \includegraphics[width=0.5\textwidth]{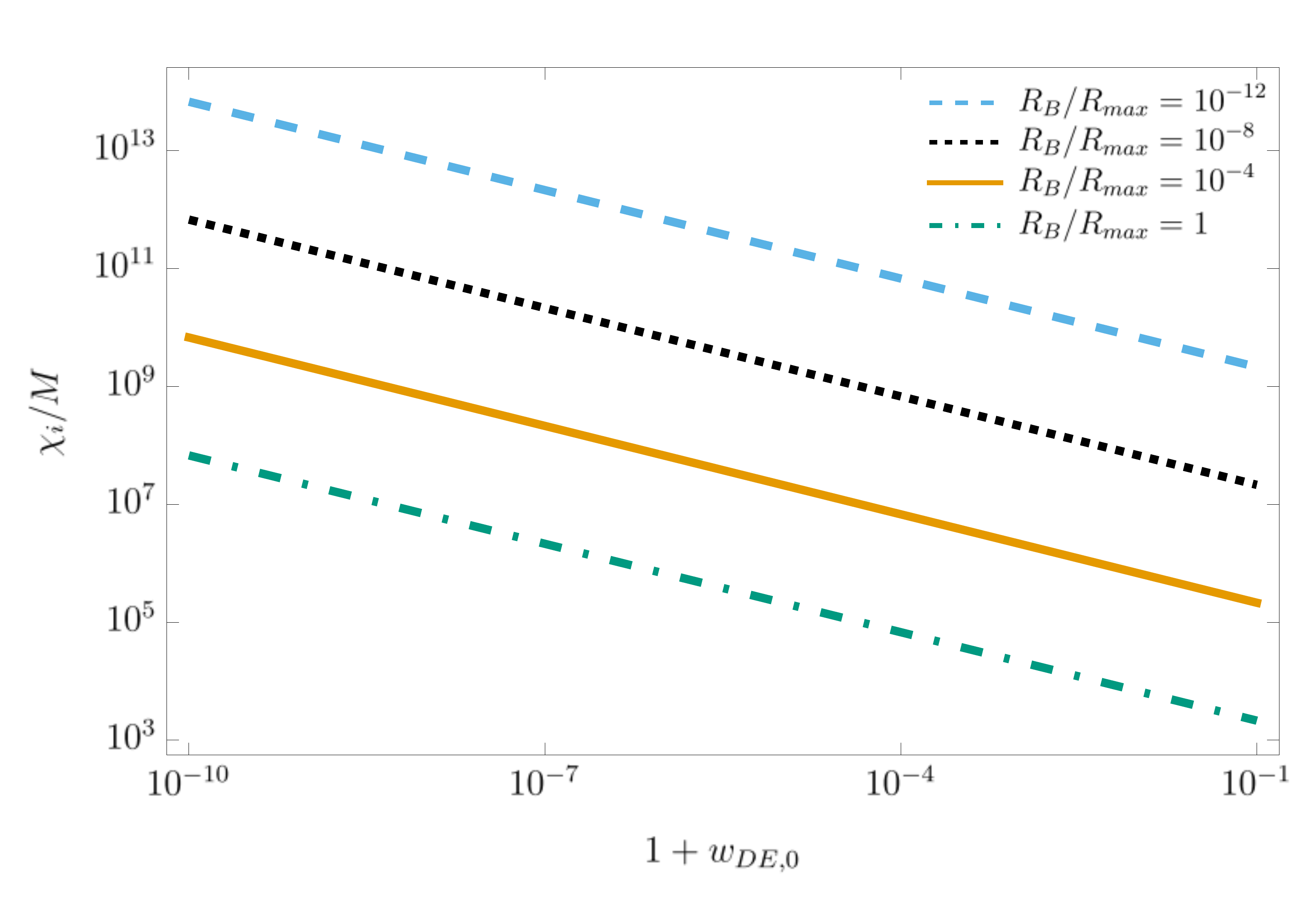}
    \caption{Parameter space demonstrating relationship between $\chi_i/M$ and dark energy equation of state $w_{DE,0}$ for various initial fractions $R_B$ of $B$-radiation. For all dark energy scenarios $\chi_i/M \gg 1$, though obtaining $\chi_i/M \gtrsim 10^{15}$ requires both extreme dark energy equation of state and a very small value of $R_B$. }
    \label{fig:RB_chiM}
\end{figure}

\subsection{Inflation \label{app:INFLdeviations}}
The arguments for an inflationary scenario are simpler because the deviation contributes strictly decaying energy densities and thus is never significant. In an inflating background with $w\simeq -1$ and $a \propto \tau^{-1}$, $\delchi$ evolves according to 
\beq
\delchi \prpr -\frac{2}{\btau} \delchi \pr + \frac{a_i^2 m^2 \tau_i^2}{\eps}\btau^2 \delchi \simeq 28 \delchi_i \btau^{-6}.
\eeq
This has the solution $\delchi = \delchi_{h}+ \delchi_{in}$, with the solution to the homogeneous equation as
\begin{align}
\delchi_h = &C_1 a^{-5/2} j_{-5/4}\left( \frac{m/\sqrt{u_i}}{2H_i} (a/a_i)^{-2} \right) +  C_2 a^{-5/2} j_{1/4}\left( \frac{m/\sqrt{u_i}}{2H_i} (a/a_i)^{-2} \right),
\end{align}
and the solution to the inhomogeneous equation as
\begin{align}
\delchi_{in} = &\kappa \pi \delchi_i \biggl[  -\lp \frac{64}{21 \Gamma(-7/4)}\rp x^{-11/4} J_{3/4}(\frac{x}{2})_{1}F_{2}\left(\{-\frac{7}{4}\}; \{-\frac{3}{4},\frac{1}{4} \}; -\frac{x^2}{16} \right)\nonumber\\
&+ \frac{7}{8}  x^{-5/4}J_{-3/4}(\frac{x}{2}) G^{2,0}_{1,3}\left( \{2 \};\{0,1,-\frac{3}{4} \};\frac{x^2}{16} \right)\biggr], \label{delchininf}
\end{align}
where $x\equiv \kappa^{1/2} \btau^2$, $_mF_n (z)$ is the generalized hypergeometric function, and $G^{m,n}_{p,q}(z)$ is the Meijer $G$ function. During inflation $H \simeq \mathrm{const}$, so $x = (m/H_i)(u)^{-1/2}$. The constraint Eq.~(\ref{eqn:mHKEcond}), evaluated at the end of inflation, gives $x \gtrsim (4 u_E)^{1/2}$, where $u_{E} \simeq 1$ corresponds to the end of inflation. Hence $x$ decays from approximately $ x_i \simeq 4/u_i^{1/2} \sim 4 e^{2N}$ to $ x_E \sim 4$ during inflation, so we can safely take the $x \gg 1$ limit when evaluating these expressions. The solution to the homogeneous equation can be written in terms of $x$ as
\begin{align}
\delchi_h = &C_1 x^{5/4} j_{-5/4}\left(\frac{x}{2}\right) +   C_2 x^{5/4} j_{1/4}\left(\frac{x}{2} \right).
\end{align}
Asymptotically, $j_n(x) \simeq x^{-1}\sin(x - n\pi/2)$ for $x\gg1$ and $x \gg n(n+1)/2$, which gives 
\begin{align}
\delchi_h = & C_1 x^{1/4} \sin(x+5\pi/8) + C_2x^{1/4} \sin(x-\pi/8).
\end{align}
The potential energy contributed by these terms is initially subdominant and subsequently decays enormously, by roughly $e^{N/2}$, so it remains negligible. For the kinetic energy, using $d/d\tau = 2 a_i H_i \kappa^{1/4} x^{1/2} d/dx$, the leading term for the kinetic energy $\delchi_h^{\prime 2}/a^2$ comes from the derivative hitting the oscillating terms. This yields  contributions that go as as $x^{5/2}$, i.e. as $a^{-5}$, and so the kinetic contribution of the oscillating terms is also decaying and therefore will not spoil cosmic acceleration.  Examining $\delchi_{in}$ in the $x\gg 1$ limit gives
\begin{align}
   \delchi_{in}/\delchi_i &\simeq \frac{\kappa \sqrt{\pi}}{48}\Gamma(5/4)x^{-3/4}\lb 5 \cos \lp x/2 - \pi/8 \rp +16 x\sin \lp x/2 - \pi/8  \rp \rb  \\  
   \delchi_{in}\pr /\delchi_i &\simeq \frac{a_i H_i \kappa^{5/4} \sqrt{\pi}}{1536} \Gamma(5/4) x^{-5/4}\lb (512 x^2 -105) \cos \lp x/2 - \pi/8 \rp +96 x\sin \lp x/2 - \pi/8  \rp \rb .
\end{align}
The dominant term for $\delchi_{in}$ goes as $x^{1/4}$, and hence $a^{-1/2}$. Similarly, the leading kinetic term goes as $x^{3/2}/a^{2}$, and hence, $a^{-5}$. Because $\delchi$ contributes decaying energy density, it can always be neglected, it remains subdominant to both $\chi_c$ and the $E$ vev, and it does not spoil inflation. We confirm these analytic conclusions with numerical results in Figure \ref{fig:infrho}. We note these conclusions do not change for non-zero $E_i$, as the leading behavior of the dominant terms is the same, i.e. the energy density of $\delchi$ is strictly decaying. The only caveat is that $\rho_{E \delchi}$ cannot dominate the initial energy density, which requires $j \lesssim u_i^{1/2}$. 

\begin{figure}[h!]
\includegraphics[width=1\textwidth]{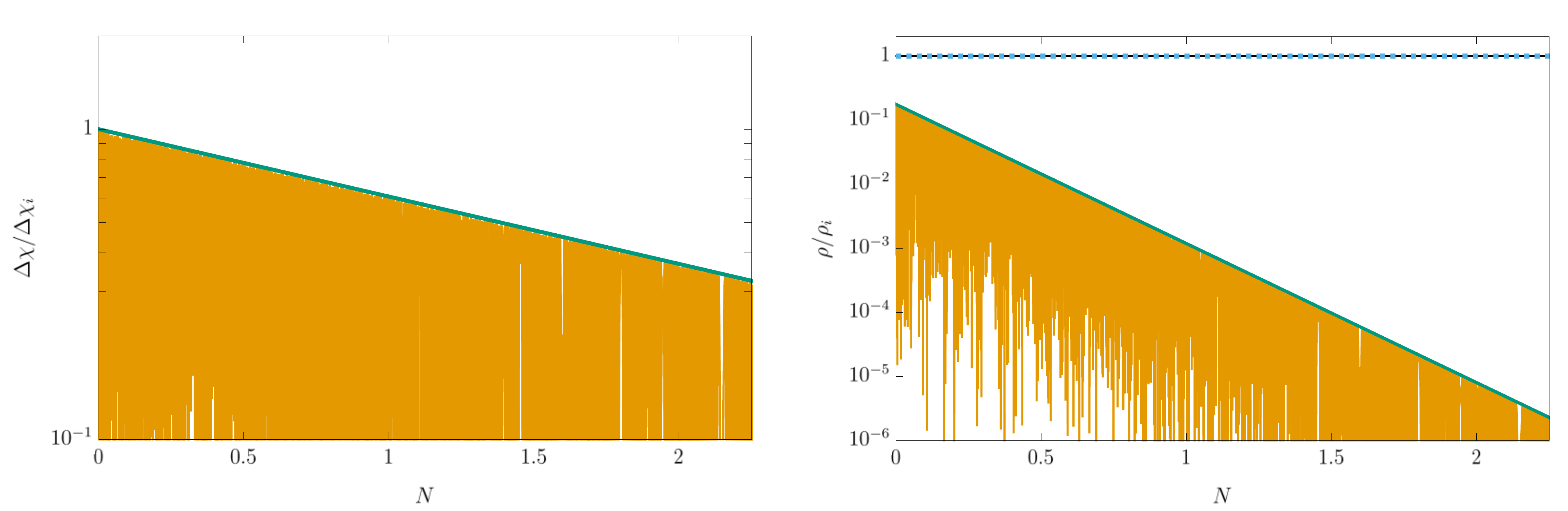}
\caption{Left panel: numerical evolution of $\delchi$ (orange, solid) over several efoldings $N$ of inflation, demonstrating $a^{-1/2}$ decay (green). Note the oscillations are fast, so the orange lines appear dense. Right panel: Comparison of numerically computed energy densities, in units of initial energy density $\rho_i$, over the same period. The total energy density (light blue, dot-dashed) and $V_{\chi_c}$ (black, dotted) are nearly constant and equal while  $\rho_{\delchi}$ (orange, solid, amplified by $10^{100}$) decays as $a^{-5}$ (green, solid).}
\label{fig:infrho}
\end{figure}

\end{widetext}
\vfill
\eject

\bibliography{main}

\end{document}